\documentclass[doublecol]{epl2} 

\usepackage{bm}
\usepackage{amsmath}

\title{Vortex-boson duality in four space-time dimensions}
\author{M. Franz}

\institute{                    
Department of Physics and Astronomy,
University of British Columbia, Vancouver, BC, Canada V6T 1Z1
}

\abstract{A continuum version of the vortex-boson duality in (3+1) dimensions is 
formulated and its implications studied in the context of a pair 
Wigner crystal in underdoped cuprate superconductors.
The dual theory to a phase fluctuating superconductor (or superfluid) 
is shown to be a theory of bosonic strings interacting through a Kalb-Ramond
rank-2 tensorial gauge field. String condensation produces Higgs mass for the
gauge field and the expected Wigner crystal emerges as an interesting 
space-time  analog of the Abrikosov lattice.}

\begin{document}
\newcommand{\bk}{{\bf k}}
\newcommand{\bp}{{\bf p}}
\newcommand{\bv}{{\bf v}}
\newcommand{\bx}{{\bf x}}
\newcommand{\bX}{{\bf X}}
\newcommand{\tbq}{\tilde{\bf q}}
\newcommand{\tq}{\tilde{q}}
\newcommand{\bQ}{{\bf Q}}
\newcommand{\br}{{\bf r}}
\newcommand{\bR}{{\bf R}}
\newcommand{\bB}{{\bf B}}
\newcommand{\bA}{{\bf A}}
\newcommand{\bK}{{\bf K}}
\newcommand{\Bomega}{{\bf \Omega}}
\newcommand{\cS}{{\cal S}}

\maketitle

When particles in quantum many body 
systems interact very strongly standard perturbation
techniques break down and, in dimensions greater than 1, {\em dualities} 
often provide the only insights into the physics of such systems.
Duality transformations, in general, map the strongly coupled sector of one
theory onto the weakly coupled sector of another. The original Kramers-Wanier
duality \cite{kramers1} for the Ising ferromagnet represents a prime example of 
such a mapping. Dualities permeate modern statistical, condensed matter
and particle physics, and have emerged recently as a key tool in string
theory. 

In condensed matter physics perhaps the most useful and influential 
duality is the one connecting vortices and bosons in two spatial dimensions
\cite{dasgupta,nelson,lee_fisher}. This duality, hereafter referred to as 
Lee-Fisher duality,  maps the system of interacting
bosons  in (2+1)D onto a fictitious superconductor in an external magnetic field
whose flux in the temporal direction is proportional to the density of the 
original bosons. It shows that Mott insulator, proximate to the phase 
fluctuating boson condensate, can be viewed as the
Abrikosov vortex lattice of the dual superconductor. This deep connection 
has been exploited in modeling systems ranging from quantum spins to
fractional quantum Hall effect, and most recently cuprate superconductors.

In cuprates such considerations are motivated by the experimental
findings of static checkerboard patterns in the charge density of very
underdoped samples \cite{vershinin1,hanaguri1,hashimoto1} which have 
been interpreted as evidence for a Cooper pair Wigner crystal (PWC)
\cite{chen1}. The latter can be most naturally understood by appealing
to the Lee-Fisher duality \cite{tesanovic1,anderson1,balents1}.
However, recent analysis of the vibrational modes
of such a PWC  \cite{tami1,taillefer1} indicates that it is 3-dimensional 
(in the sense that vibrations propagate in all 3 
space dimensions) and it is thus unclear how the inherently two-dimensional
Lee-Fisher duality applies to this situation.
The problem can be stated as follows. A key role in the formulation
of  the Lee-Fisher duality is
played by vortices which appear (in pairs of opposite vorticity)  
near the transition to the Mott insulating 
phase as quantum fluctuations of the 
system. The dual relationship between bosons and vortices however
exists only in two spatial dimensions where the latter can be regarded 
as {\em point particles}. 
In three space dimensions vortices form oriented loops and can no longer 
be thought of as particles.  
The question thus arises how to understand the formation of a PWC in the
three dimensional phase fluctuating superconductor indicated by experiments
\cite{vershinin1,hanaguri1,hashimoto1}.

In this Letter we point a way out of this conundrum by constructing a
(3+1) dimensional implementation of vortex-boson duality using a 
representation of vortex loops as relativistic bosonic strings. We then
show that string condensation indeed produces a ground state that can be 
characterized as an insulating crystal of Cooper pairs and discuss some of 
its unique properties.

We remark that the lattice formulation of such (3+1)D duality has been given
long time ago \cite{peskin1,savit1} and was used recently to study
exotic fractionalized phases \cite{wen1,senthil1} in (3+1)D. Here, by contrast, 
we formulate a continuum version which shows that a boson (Cooper pair) 
crystal can emerge from a phase fluctuating superfluid (superconductor) even
in the absence of any underlying lattice structure. This is exactly the limit 
apparently relevant to cuprates \cite{tami1}.  On the formal side this 
continuum approach also reveals an intimate connection to the theory of 
bosonic strings and enables us to employ in our calculations some of the 
string theory technology. 

We begin by considering a continuum theory of a superconductor in (3+1)
space-time dimensions defined by the Euclidean 
partition function $Z=\int{\cal D}[\Psi,\Psi^*]\exp(
-\int_0^{\beta} d\tau\int d^3 x {\cal L})$ with $\Psi=|\Psi|e^{i\theta}$ a
scalar order parameter and the Lagrangian density  
\begin{eqnarray}\label{L0}
{\cal L} = {1\over 2}\tilde K\left|\left(\partial_\mu - 
{2ie}A_\mu\right)\Psi\right|^2  + U(|\Psi|^2).
\end{eqnarray}
The Greek index $\mu=0,1,2,3$
labels the temporal and spatial components of (3+1) dimensional vectors,
and we use natural units with $\hbar=c=1$. $U$ is a potential
function that sets the value of the order parameter $\Psi$ in the 
superconducting state in the absence of fluctuations.
The electromagnetic vector potential $A$ is explicitly displayed in order
to track the charge content of various fields. If we allowed $A$ to fluctuate
then Eq.\ (\ref{L0}) would coincide with the well known Abelian Higgs model.

We now focus on the fluctuations in the phase $\theta$ by fixing the amplitude
$|\Psi|=\Psi_0$ at the minimum of $U$,
\begin{eqnarray}
{\cal L} = {1 \over 2} K 
\left(\partial_\mu \theta - {2e}A_\mu\right)^2,
\end{eqnarray} 
where $K=\tilde{K}\Psi_0^2$ represents the phase stiffness. The first few 
steps of the duality mapping proceed just as in (2+1)D. We first 
decouple the quadratic term with a real auxiliary field, $W_\mu$, using the
familiar Hubbard-Stratonovich transformation, obtaining
\begin{equation}\label{eq:W}
{\cal L} = {1 \over 2K}W^2_\mu +iW_\mu(\partial_\mu\Theta - {2e}A_\mu)+
iW_\mu(\partial_\mu\theta_s).
\end{equation}
We have also decomposed the phase into a 
smooth part $\theta_s$ and singular part $\Theta$ containing vortex
lines. 

Gaussian integration over $\theta_s$ leads to a constraint
\begin{equation}\label{con}
\partial_\mu W_\mu=0, 
\end{equation}
which reflects conservation of electric charge. In (2+1)D one enforces this 
constraint by expressing $W_\mu$ as a curl of an auxiliary gauge field. 
The curl operation, however, is meaningful only in 3 dimensions and herein 
lies the difficulty with higher dimensional duality. In (3+1)D we may 
enforce the constraint (\ref{con}) by writing
\begin{equation}\label{res}
W_\mu=\epsilon_{\mu\nu\alpha\beta}\partial_\nu B_{\alpha\beta},
\end{equation}
where $\epsilon_{\mu\nu\alpha\beta}$ is the totally antisymmetric tensor and
$B_{\alpha\beta}$ is antisymmetric rank-2 tensor gauge field.
Substituting Eq.\ (\ref{res}) back into the Lagrangian and performing 
integration by parts in the term containing $\Theta$ we obtain
\begin{equation}\label{L2}
{\cal L} = {H^2_{\alpha\beta\gamma} \over 3K}
-iB_{\alpha\beta}(\epsilon_{\alpha\beta\mu\nu}\partial_\mu\partial_\nu\Theta)
-i2e(\epsilon_{\mu\nu\alpha\beta}\partial_\nu B_{\alpha\beta})A_\mu,
\end{equation}
where $H_{\alpha\beta\gamma}=\partial_\alpha B_{\beta\gamma}
+\partial_\beta B_{\gamma\alpha}+\partial_\gamma B_{\alpha\beta}$ is the 
tensorial field strength which should be thought of as a generalization
of the Maxwell field strength $F_{\mu\nu}$.

The above Lagrangian exhibits several notable features. First, it 
possesses invariance under the gauge transformation
\begin{equation}\label{gauge}
B_{\alpha\beta} \to B_{\alpha\beta}+\partial_{[\alpha}\Lambda_{\beta]}
\end{equation}
for an arbitrary smooth vector function $\Lambda_\mu$. The square brackets 
represent antisymmetrization, e.g.\ $\partial_{[\alpha}\Lambda_{\beta]}=
\partial_{\alpha}\Lambda_{\beta}-\partial_{\beta}\Lambda_{\alpha}$. This
gauge invariance reflects conservation of vorticity 
in the original model. Second, from the last term in $\cal L$ we may
immediately deduce that the electric four-current is related to $B$ by
$j_\mu=2e(\epsilon_{\mu\nu\alpha\beta}\partial_\nu B_{\alpha\beta})$.
The charge density, in particular, can be written as 
\begin{equation}\label{e}
\rho=j_0=2e(\epsilon_{ijk}\partial_iB_{jk}),
\end{equation}
where Roman indices run over spatial components only.

The second term in $\cal L$ informs us that field $B$ is minimally
coupled to the ``vortex loop current''
\begin{equation}\label{vor}
\sigma_{\alpha\beta}(x)=
\epsilon_{\alpha\beta\mu\nu}\partial_\mu\partial_\nu\Theta(x),
\end{equation}
which is an antisymmetric rank-2 tensor quantity. For a smooth function the
right hand side of Eq.\ (\ref{vor}) would vanish since the derivatives would
commute. In 3 spatial dimensions, single valuedness of $e^{i\Theta(\bx)}$ 
permits {\em line singularities} in $\Theta(\bx)$ such that it varies
by an integer multiple of $2\pi$ along any line that encircles the singularity.
These are the vortex loops. 

To implement the duality transformation we now shift our point
of view from the phase field $\Theta(x)$ to the vortex loop {\em worldsheets}. 
These describe the evolution of vortex loops in imaginary time and should be 
thought of in analogy with worldlines of point particles. The worldsheets
are specified by a set of 2-parameter vector functions 
$X_{\mu}^{(n)}(\sigma_1,\sigma_2)$ where $n$ labels the individual loops.
We take $\sigma_1$ to be time-like, and correspondingly vary between
0 and the inverse temperature $\beta$, and $\sigma_2$ space-like, which by 
convention varies from 0 to $2\pi$ for closed loops. (Since vortices can only
terminate on magnetic monopoles we consider only closed vortex loops here.)
Clearly, given a set of worldsheets $X_{\mu}^{(n)}$ one can 
reconstruct the phase field $\Theta(x)$ up to any smooth contribution.

A surface element of a worldsheet is characterized by a rank-2 antisymmetric
tensor 
\begin{equation}\label{sig}
\Sigma_{\mu\nu}^{(n)}={\partial X_{[\mu}^{(n)}\over \partial\sigma_1}
{\partial X_{\nu]}^{(n)}\over \partial\sigma_2}.
\end{equation}
It is straightforward to show that the loop current (\ref{vor}) is related 
to the worldsheet by
\begin{equation}\label{sig2}
\sigma_{\mu\nu}(x)=2\pi\sum_n\int d^2\sigma \Sigma_{\mu\nu}^{(n)}
\delta\left( X^{(n)}-x\right).
\end{equation}
This relation allows us to rewrite the partition function as a functional
integral over the vortex loop worldsheets $X_{\mu}^{(n)}$. 
We thus have $Z=\int{\cal D}[X] \exp(-{\cal S})$ with
\begin{eqnarray}\label{action}
{\cal S}&=&\sum_n \int d^2\sigma \left[{\cal T}\sqrt{
\Sigma_{\mu\nu}^{(n)}\Sigma_{\mu\nu}^{(n)}}
-2\pi i\Sigma_{\mu\nu}^{(n)}B_{\mu\nu}(X^{(n)})
\right]\nonumber \\
&+&{1\over 3K}\int d^4x H_{\alpha\beta\gamma}^2 +{\cal S}_{\rm int}+
{\cal S}_{\rm  Jac}.
\end{eqnarray}
We recognize the first line as the celebrated Nambu-Goto action
\cite{nambu-goto} for 
bosonic strings propagating in the presence of a background Kalb-Ramond
gauge field $B_{\mu\nu}$ \cite{kalb1}. The first term can be interpreted as the
reparametrization invariant surface area of the string worldsheet with 
the string tension ${\cal T}$. Although such term does not appear
explicitly in Eq.\ (\ref{L2}) it would arise in a more careful treatment of
our starting theory (\ref{L0}) had we retained the cost of the suppression
of the order parameter amplitude $|\Psi|$ near the vortex core. 
The second and the third terms follow directly from Eq.\ (\ref{L2}) and
describe long range interactions between strings mediated by the superflow,
now represented by the Kalb-Ramond gauge field.

${\cal S}_{\rm int}$ contains short range interactions between strings that 
would also arise from a more careful treatment of the core physics. Finally,
${\cal S}_{\rm  Jac}$ represents the Jacobian of the transformation from
phase variable $\Theta$ to string worldsheets $X_{\mu}^{(n)}$. This last term
plays an important role in the quantization of our string theory. It is well known
that a fundamental string can be consistently quantized only in the critical
dimension, which for bosonic string is $D=26$ \cite{crit}. A question then 
arises as to how we quantize our vortex strings in (3+1)D; after all we 
started 
from a well defined field theory (\ref{L0}) and  we expect the string theory 
(\ref{action}) derived from it to also be well behaved. The answer lies in 
the fact that our strings are {\em not} fundamental; rather they are
Nielsen-Olesen type strings \cite{nielsen1} with intrinsic
thickness defined by the core size. It was shown by Polchinski and
Strominger \cite{pol1} that terms in  ${\cal S}_{\rm  Jac}$, which would be 
absent in the case of a fundamental string, precisely cancel the conformal 
anomaly responsible for the high critical dimension. Vortex strings are indeed 
well behaved in the physical dimension.

We are now ready to complete the duality mapping. Our main goal will be 
to understand the string-condensed phase, analogous to the vortex-condensed 
phase in the (2+1)D Lee-Fisher duality.
To this end we must pass to second quantized string theory, a ``string-field
theory''. This can be done rigorously in the so called light-cone gauge
\cite{kaku1} or using the Becchi-Rouet-Stora-Tyutin (BRST) procedure 
\cite{brst}. Here we opt for a less rigorous but physically much more 
transparent procedure devised in Ref.\ \cite{rey1} which provides a 
straightforward route towards the description of the string-condensed phase.

The central concept in the string-field description is the wave {\em functional}
$\Phi[X]$, a complex-valued functional defined on the space of one-parameter
string trajectories $\{X_\mu(\sigma_2),\sigma_2=(0,2\pi)\}$, which we regard
as cross sections of the string worldsheet $X_\mu(\sigma_1,\sigma_2)$
at fixed value of $\sigma_1$. The physical significance of $\Phi[X]$ is 
most readily visualized in the so called static parametrization: in a 
chosen Lorentz
frame of reference take $X_0=\sigma_1=\tau$ and $\bX=\bX(\tau,\sigma)$,
with $\tau$ the imaginary time. $\Phi[\tau,\bX]$ then represents the 
quantum mechanical amplitude for finding the string in configuration 
$\{\bX(\sigma),\sigma=(0,2\pi)\}$ at time $\tau$. 

The second
quantized action for the string functional takes the form \cite{rey1}
\begin{eqnarray}\label{action2}
{\cal S}&=& \int{\cal D}[X]\int d\sigma \sqrt{h}
\bigl[\big|(\delta/\delta\Sigma_{\mu\nu}-2\pi iB_{\mu\nu})\Phi[X]\big|^2
\nonumber \\
&+&{\cal M}_{\rm eff}^2\big|\Phi[X]\big|^2\bigr]
+{1\over 3K}\int d^4x H_{\alpha\beta\gamma}^2 +{\cal S}'_{\rm int}.
\end{eqnarray}
The plaquette derivative $\delta/\delta\Sigma_{\mu\nu}$ quantifies the 
variation of the functional $\Phi[X]$ upon modifying the path $X$ by an 
infinitesimal loop $\Delta X$  which sweeps the surface element 
$\delta\Sigma_{\mu\nu}$. The loop space metric 
$h=(\partial X_\mu(\sigma)/\partial\sigma)^2$ is needed to preserve the
reparametrization invariance of the action.
${\cal S}'_{\rm int}$ represents short range string interactions and 
contains terms cubic and higher order in $|\Phi|$. All contributions 
quadratic in $|\Phi|$ have been folded into the effective string mass 
${\cal M}_{\rm eff}$. The action (\ref{action2}) remains invariant
under the gauge transformation (\ref{gauge}) if we require $\Phi$ to transform
as
\begin{equation}\label{gauge2} 
\Phi[X]\to \Phi[X]e^{-4\pi i\int dX_\mu\Lambda_\mu}.
\end{equation}

String condensation occurs when ${\cal M}_{\rm eff}^2$ becomes negative. 
The string functional then develops nonzero vacuum expectation value, 
$\langle 0| \Phi[X] |0\rangle \neq 0$. The simplest case is that of a 
uniform string condensate, 
\begin{equation}\label{unif} 
\langle 0| \Phi[X] |0\rangle=\Phi_0={\rm const}. 
\end{equation}
Physically, this simply means that {\em any} string configuration is 
equally probable. This ansatz, however, cannot describe a phase
disordered superconductor. To see this note that substituting Eq.\ (\ref{unif})
into action (\ref{action2}) produces a mass term for the Kalb-Ramond gauge
field. Such a mass term then leads to the Meissner effect: the gauge 
field is expelled from the interior of the sample, $B_{\mu\nu}=0$.
In view of Eq.\ (\ref{e}), this corresponds to complete expulsion of charge
from the system, which is not the situation we are interested in.

What we seek is the analog of the Abrikosov vortex state in which
the field can penetrate in quantized increments. 
We thus consider a more general ansatz, which allows both the amplitude and 
the phase of $\Phi[X]$ to vary:   
\begin{equation}\label{abr} 
\langle 0| \Phi[X] |0\rangle=\Phi_0 e^{\int d\sigma
[\zeta\sqrt{X'^2}\ln f(X)+2\pi iX'_\mu\cdot\Omega_\mu(X)]
}.
\end{equation}
Here $X=X(\sigma)$,  $X'=\partial_\sigma X(\sigma)$, $f(x)$ and  
$\Omega_\mu(x)$ are real scalar and vector functions 
parametrizing the functional, and $\zeta$ is a parameter with the dimension of
inverse length. $f(x)$ is nonnegative and should be thought of as the 
space-time-dependent amplitude of the string condensate. Specifically, 
$f=1$ corresponds to uniform condensate amplitude while $f>1$ ($f<1$)
describes its local enhancement (depletion).
$\Omega_\mu(x)$ determines the phase of the string condensate. Substituting
Eq.\ (\ref{abr}) to (\ref{action2}) we obtain
${\cal S}=\int d^4x{\cal L}$ with 
\begin{eqnarray}\label{action3}
{\cal L}&=& {\Phi_0^2\over 2}
\left[\pi^2f^2(\partial_{[\mu}\Omega_{\nu]}-2B_{\mu\nu})^2
+\zeta^2(\partial_\mu f)^2  +{\cal V}(f^2)\right] \nonumber \\
&+&
{1\over 3K} H_{\alpha\beta\gamma}^2.
\end{eqnarray}
We observe that  any smooth part of $\Omega$ can be eliminated by 
the gauge transformation (\ref{gauge2}). Thus, only the singular part of  
$\Omega$ has physical significance. Indeed we note that it is 
permissible for $\Omega$ to be {\em multiply valued} as long as the wave
functional (\ref{abr}) remains single valued.
A configuration of specific interest to us contains {\em monopoles}
in the spatial part of $\Omega=(\Omega_0,\Bomega)$, 
\begin{equation}\label{mon}
\nabla\cdot(\nabla\times\Bomega)=\sum_a Q_a\delta^{(3)}(\bx-\bx_a),
\end{equation}
where $\bx_a$ and $Q_a$ label the position and the charge of the 
$a$-th monopole. 
Single valuedness of (\ref{abr}) demands that  $Q_a$ be 
{\em integer}. We shall see that such 
singularities represent sources for $B_{\mu\nu}$, just as 
vortices in a superconductor act as sources for the magnetic field. 

We now analyze the action (\ref{action3}) in the presence of static monopole 
configurations in $\Bomega$. To this end we adopt a dual mean-field 
approximation (DMFA) which neglects quantum fluctuations of all the fields. We
emphasize that in terms of the original phase degrees of freedom, DMFA
describes a highly nontrivial quantum fluctuating state. In addition, we
perform a dual ``London'' approximation, $f(\bx)=1$, which should be 
adequate as long as the monopoles are relatively dilute.  (This approximation
fails in the small region near the monopole center where $f\to 0$.)
The ground state energy of the system can then 
be written as
\begin{equation}\label{ge}
{\cal E}= {1\over 2}\int d^3x
\left[\pi^2\Phi_0^2(\partial_{[i}\Omega_{j]}-2B_{ij})^2
+{1\over K}(\epsilon_{ijk}\partial_i B_{jk})^2
\right].
\end{equation}
Minimizing with respect to $B_{ij}$ leads to the Euler-Lagrange equation
\begin{equation}\label{el}
\pi^2\Phi_0^2(2B_{ij}-\partial_{[i}\Omega_{j]})
-{1\over 2eK}\epsilon_{ijk}\partial_k\rho=0,
\end{equation}
where we used Eq.\ (\ref{e}). Next, acting on all terms by 
$\epsilon_{ijl}\partial_l$ and defining a dual ``penetration depth''
$\lambda_d^{-2}=2\pi^2\Phi_0^2K$,
we obtain an equation for charge density $\rho(\bx)$
\begin{equation}\label{el1}
\rho-\lambda_d^2\nabla^2\rho=2e\nabla\cdot(\nabla\times\Bomega).
\end{equation}
This equation resembles the London 
equation for the $z$-component of magnetic field in the presence of an
Abrikosov lattice of vortices and can be analyzed by similar methods. 
The key difference is that, in light of 
Eq.\ (\ref{mon}), the right hand side describes a collection of {\em 
point sources} in three space dimensions whereas Abrikosov vortices
are line singularities described by $\delta^{(2)}$. Below we briefly
summarize some main results of this analysis and the relevant
details will be given elsewhere \cite{franz00}.

Eq.\ (\ref{el1}) can be solved for an arbitrary arrangement
of monopole positions and charges to obtain
\begin{equation}\label{rho1}
\rho(\bx)=2e\sum_a Q_a {e^{-|\bx-\bx_a|/\lambda_d}\over 4\pi\lambda_d^2 
|\bx-\bx_a|}.
\end{equation}
It is easy to show that the total electric charge associated with a monopole
is $2eQ_a$; the charge is quantized in the units of $2e$, as expected.
At finite charge density, monopoles with like charges repel by Yukawa
potential $\sim e^{-r/\lambda_d}/r$ and 
the ground state is a Bravais lattice of elementary ($Q_a=1$) monopoles.
This leads to periodic modulation in $\rho(\bx)$ with charge $2e$ per unit 
cell: a pair Wigner crystal in three space dimensions. 

An appealing overall picture thus emerges. Vortex loops in a
(3+1)-dimensional superconductor (or superfluid) can be efficiently described
as bosonic strings interacting through a rank-2 tensorial Kalb-Ramond gauge
field $B_{\mu\nu}$. In the non-superconducting phase, strings proliferate and
condense, producing Higgs mass for the gauge 
field. In the Higgs phase the only way for $B_{\mu\nu}$ to penetrate into the 
bulk of the system is to set up quantized monopole-like singularities in the 
phase
$\Omega_\mu$ of the string condensate wave functional. These singularities 
then act as point sources for $B_{\mu\nu}$. The associated electric charge 
density $\rho$, which is closely related to the Kalb-Ramond field strength
$H_{\mu\nu\lambda}$, is then governed by a London-like equation (\ref{el1}). 
For a periodic array
of point sources, such as will form at finite charge density, a 3-dimensional 
pair Wigner crystal emerges with charge distribution given by Eq.\ 
(\ref{rho1}).

The duality discussed above establishes vortex-loop condensation as a
concrete mechanism for the formation of a pair Wigner crystal in a 
3-dimensional quantum phase fluctuating superconductor. It explains how a 
3d PWC can form in underdoped cuprates and allows for detailed computations
of its structure and vibrational modes \cite{franz00} which are of direct 
experimental interest.

The author is indebted to T. Davis,  M.P.A. Fisher, S.-S. Lee, T. Pereg-Barnea,
C. Weeks
and Z. Te\v{s}anovi\'c for stimulating discussions and correspondence. This 
work was supported by NSERC, CIAR and the A.P. Sloan Foundation.

\end{document}